\documentclass[pra,
                preprint,
                amsmath,
                amssymb,
                showpacs,
                superscriptaddress]{revtex4}
\usepackage{graphicx} % Include figure files
\usepackage{dcolumn}  % Align table columns on decimal point
\usepackage{bm}       % bold math
\usepackage{amsfonts}
\usepackage{dsfont}
\usepackage{csquotes}

\usepackage{ulem}
\usepackage{color}

\begin{document}

\title{Influence of Vacancy Defect in Solid High-Order Harmonic Generation}

\author{Adhip Pattanayak}
\affiliation{%
Department of Physics, Indian Institute of Technology Bombay, Powai, Mumbai 400076, India }

\author{Mrudul M. S.}
\affiliation{%
Department of Physics, Indian Institute of Technology Bombay,
           Powai, Mumbai 400076, India }

\author{Gopal Dixit}
\email[]{gdixit@phy.iitb.ac.in}
\affiliation{%
Department of Physics, Indian Institute of Technology Bombay, Powai, Mumbai 400076, India }

\date{\today}

%%%%%%%%%%%%%%%%% END OF PREAMBLE %%%%%%%%%%%%%%%%
\begin{abstract}
The present work discusses the impact of vacancy defects in solid high-order harmonic generation. 
The total energy cut off of the high-order harmonic  spectrum  increases as a function of  
concentration of vacancy defect, and the total spectrum gradually turns into a single slanted 
spectrum without having an abrupt transition between primary and secondary  plateaus. 
The spectral intensity of the below band-gap harmonics in a solid with vacancy defects 
is enhanced significantly in comparison to the harmonics in a pristine solid. The changes in the 
harmonic spectra are understood in terms of their effective band structures. 
The presence of vacancy defects  
breaks the translational symmetry of the unit cell locally.  
As a consequence of this, new defect states appear, which open additional paths for the electron dynamics. 
The ill-resolved electron trajectories in the Gabor profile confirm the interference of 
additional paths. Moreover, the single slanted high-order harmonic spectrum carries a unique signature of vacancy defects in comparison to the high-order 
harmonic spectrum corresponding to 
solids with defects such as underdoping or overdoping. 
\end{abstract}

\maketitle 

\section{Introduction}

High-order harmonic generation (HHG) is a process in which 
a strong laser pulse interacts with gaseous atoms or molecules 
and  electromagnetic radiation, consisting of higher-order harmonics of the driving laser pulse, is emitted as an outcome of this interaction~\cite{ferray1988multiple}.  
The underlying physical mechanism of HHG is well understood by 
the semiclassical three-step model~\cite{corkum1993plasma}. 
HHG in a gaseous medium offers the route to generate 
attosecond pulses ~\cite{bartels2002generation, sansone2006isolated, paul2001observation} 
as well as to interrogate electron and nuclear dynamics in atoms and molecules
on their natural timescales~\cite{krausz2009attosecond, smirnova2009high, dixit2012, bredtmann2014x, bucksbaum2007, corkum2007, lepine2014attosecond}. 

Recently, a variety of solid-state systems such as dielectrics, semiconductors, nano-structures  
and noble gas solids 
have been used to generate strong-field driven 
high-order harmonics of terahertz and mid-infrared 
driving frequencies~\cite{ghimire2011observation, ghimire2011redshift, zaks2012experimental, schubert2014sub, vampa2015all, vampa2015linking, hohenleutner2015real, luu2015extreme, ndabashimiye2016solid, you2017high, lanin2017mapping, sivis2017tailored, langer2018lightwave}.  
In the case of HHG from solids, 
the harmonic cutoff linearly depends on the strength 
and the wavelength  of the driving laser field, 
which is in contrast to HHG from gases 
where the cutoff depends quadratically on 
 the strength and the wavelength 
~\cite{ghimire2011observation, ghimire2014strong, wu2015high, du2017quasi}. 
Moreover, the harmonic spectrum from solids consists of multiple plateaus, 
which extend beyond the HHG from gases~\cite{ndabashimiye2016solid}. 
Higher electron density as well as periodicity make solids an ideal source for harmonic generation~\cite{luu2015extreme,  vampa2017merge, kruchinin2018colloquium}. 
This offers an attractive compact table-top source for  
coherent and bright attosecond pulses in extreme ultraviolet (XUV) 
and soft x-ray energy regimes. 
HHG from solids is used to image energy bands~\cite{ndabashimiye2016solid} 
and to perform tomography of impurities in solids~\cite{almalki2018high}. 
Therefore, HHG from solids opens a unique avenue to 
probe energy structures and electron dynamics in solids at their intrinsic timescales~\cite{higuchi2014strong, 
ghimire2014strong, vampa2017merge, pattanayak2019direct, kruchinin2018colloquium}.

In general, 
different forms of  defect such as vacancies, interstitials,
impurities (donor- or acceptor-type doping), etc., are inevitably present in real solids.  
Understanding how different kinds of defects affect  HHG in solids 
is an interesting and  challenging problem.

Recently, several theoretical works have discussed the influence of  defects on solid HHG  ~\cite{almalki2018high, orlando2018high, yu2018enhanced, huang2017high}.
Time-dependent density functional theory 
in one-dimension is employed to understand HHG in donor-doped, acceptor-doped and un-doped semiconductors~\cite{yu2018enhanced}. 
In donor-doped solids, the highest occupied impurity orbital found to be 
very close to the lowest conduction band.  
 The probability of electrons tunneling from the impurity  orbital to the 
 conduction band is high, as the tunneling rate depends on the energy band gap. 
In comparison to acceptor-doped and un-doped semiconductors, several orders of magnitude enhancement  in the efficiency of HHG in donor-doped semiconductor has been reported~\cite{yu2018enhanced}. 
On the other hand, Huang et al. have investigated  HHG from doped semiconductors and found that the efficiency of the second plateau from a doped semiconductor is 
enhanced by about  three orders of magnitude in comparison to an un-doped semiconductor~\cite{huang2017high}. 
The tight-binding Hamiltonian within the two-band Anderson model with disorder is employed 
to understand HHG in disordered semiconductor~\cite{orlando2018high}. 
It has been shown that the disorder is the probable cause behind the experimental observation of the well-resolved harmonic peaks in the spectrum~\cite{orlando2018high}.   
Corkum and co-workers have developed a conceptual formalism for the tomographic imaging of the shallow impurities using a one-dimensional hydrogenic model~\cite{almalki2018high}. 
All of these works have shown the possible modification of high-order harmonic spectra with different kinds of substitutional impurity, while the influence of vacancy defects on the HHG is still an open problem.
    
In the present work, we discuss  the influence of vacancy defects in solid HHG. The HHG spectra due to vacancy defects can be compared with substitutional defects such as underdoping (acceptor doping) and overdoping (donor doping). 
We will show that the harmonic spectra 
corresponding to solids with vacancy defects are significantly different from the spectra corresponding to solids with substitutional defects. 
The supercell approach is used to realize different concentrations of vacancy defects 
in solid. 
The time-dependent Schrodinger equation (TDSE) in the Bloch basis is solved to simulate the HHG
 from solids with various vacancy-defect concentrations. 
The present paper is organised as follows. In Sec. II, the theoretical model and numerical methods are presented. The results and discussion of our numerical simulations  are given in Sec. III, and the conclusion is presented in Sec. IV. 

\section{Theoretical Model}
\begin{figure}[h!]
\includegraphics[width=12 cm]{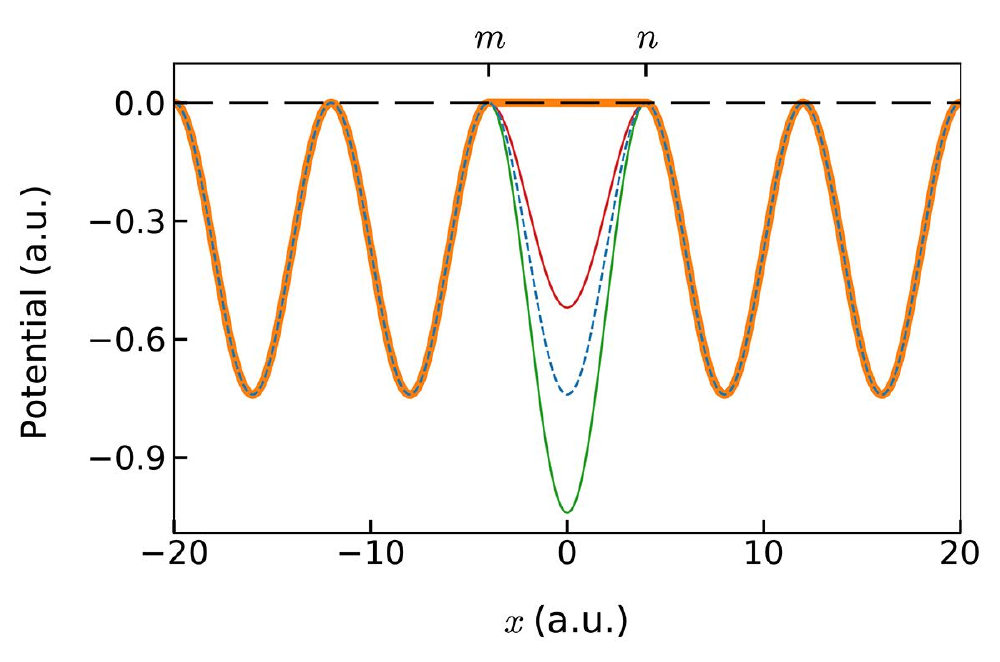}
\caption{
One-dimensional periodic potential for  a pristine (defect-free) solid  (dashed blue curve), a solid with a vacancy defect (solid orange curve), an overdoped solid (solid green curve), and an underdoped solid (solid red curve).  Here, the defect concentration is 20\% because  a single defect is present in the supercell, which is 5 times larger than the single unit cell.} \label{fig1}
\end{figure}

In this work, atomic units are used throughout unless stated otherwise. 
To obtain the high-order harmonic spectrum, the TDSE
within the single-active-electron approximation is solved
\begin{equation}\label{eq01}
i\frac{\partial }{\partial t}\left |\psi (t) \right \rangle=[\mathcal{H}_{0}+\mathcal{H}_{\textrm{int}}]\left |\psi (t) \right \rangle,
\end{equation}
where $\mathcal{H}_{0}$ is  the field-free Hamiltonian and $\mathcal{H}_{\textrm{int}}$ is the 
interaction Hamiltonian between the solid and the laser pulse. 
In the present study,  the laser pulse is considered to be linearly polarized.  
A Mathieu-type model potential is used as 
the periodic potential of the solid without defects and  is written as
\begin{equation}\label{eq02}
V(x) = -V_{0}\left(1+\cos\left(\frac{2\pi x}{a}\right)\right),
\end{equation} 
where $V_{0}$ is the depth of the potential and $a$ is the lattice constant.  
Here, $V_{0} = 0.37$ a.u. and $a = 8$ a.u. are used. 
The Mathieu-type potential is a popular potential widely used to study HHG in solids~\cite{wu2015high, liu2017wavelength, liu2017time, ikemachi2017trajectory}. 
The Mathieu-type potential is modified to incorporate the different kinds of defects 
and is written as 
\begin{equation}\label{eq03}
    \begin{split}
V(x) & = -V_{0}\left(1+\cos\left(\frac{2\pi x}{a}\right)\right) \hspace{1cm}x < m \text{  \:or\:  } x> n,\\
        & = -V_{1}\left(1+\cos\left(\frac{2\pi x}{a}\right)\right) \hspace{1cm}m \leq x \leq n.
    \end{split}
\end{equation}
The value of $V_1$ can be modified to realize different kinds of defects 
such as   vacancies ($V_1=0$), underdoping ($V_1=0.26$ a.u.) and overdoping ($V_1=0.52$ a.u.), where 
the values of $V_1$ for underdoped and overdoped  cases are adopted from Ref.~\cite{huang2017high}. 
The absence of an atom in a lattice site creates a vacancy in the periodic potential as shown in Fig.~\ref{fig1}.  
The periodic potential with a vacancy defect (solid orange curve) differs from the pristine or 
vacancy free  periodic potential (blue dashed curve) in the region between $m$ and $n$ points. 
We assume that the lattice constant $a$ is unaffected due to the presence of  a vacancy defect. 

Different concentrations of vacancy defects are achieved systematically by increasing  
the size of the supercell and keeping a single vacancy in the supercell. 
The vacancy concentration becomes $\left(\frac{100}{N}\right)\%$  when the supercell is $N$-times larger than the single unit cell.  
The  vacancy concentration becomes  4.7\%, 9.1\%, 14.2\% and 20.0\% corresponding to 
$N = 21, 11, 7$ and 5; respectively. A linearly polarized laser pulse with eight sine-square enveloped optical cycles is used to obtain the HHG spectrum. 
The wavelength and the intensity of the laser pulse are 3.2 $\mu$m and
 8.1$\times$10$^{11}$ W/cm$^{2}$, respectively.  

The HHG spectrum is simulated by solving the TDSE  
in  the velocity-gauge within the Bloch-state basis as given in 
Ref.~\cite{korbman2013quantum}. 
Brillouin zone sampling is used for pristine as well as supercell calculations.
A fourth-order Runge-Kutta method with 0.01 a.u. time-step  is 
employed to solve the coupled differential equations. 
The intensity of the high-order harmonics is obtained 
from the square of the Fourier transform of the time-dependent current as
\begin{equation}\label{eq04}
I(\omega)  = \left|~\int dt~j(t)~e^{i \omega t}~\right|^{2},
\end{equation} 
where $j(t)$ is laser-induced time-dependent current.

\section{Results and Discussion}
 
\begin{figure}[h!]
\includegraphics[scale=1.40]{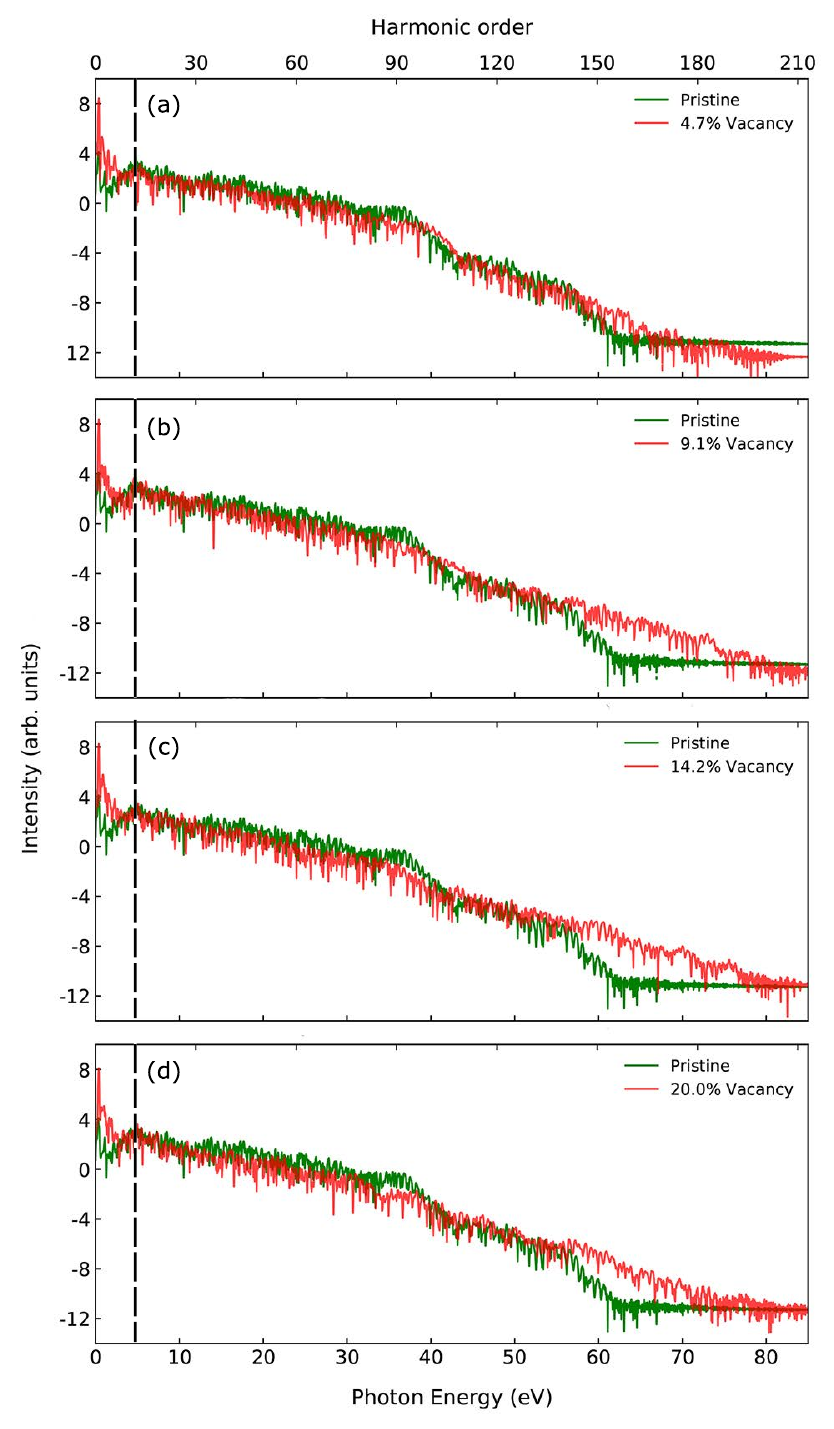}
\caption{High-order harmonic spectrum of  solids with vacancy defects of different concentrations in red color (light gray) compared to the spectrum of a  pristine solid in green color (dark gray). The spectra are for different vacancy concentrations: (a) 4.7\%, (b) 9.1\%, (c) 14.2\% and (d) 20.0\%. The minimum band gap of 4.18 eV for the  pristine solid
is marked by the dashed black vertical line.} \label{fig2}
\end{figure}  

Figure~\ref{fig2} represents the high-order harmonic spectrum of solids with vacancy defects of different concentrations (red color) compared to  the spectrum of a pristine
 solid without vacancy defects (green color). 
The  harmonic spectrum of the pristine solid reproduces unique features as it exhibits both a primary and a secondary plateau and a sudden transition from the primary plateau  to the secondary plateau
 with clear cutoffs.  
The primary plateau lies between the 12th and 90th harmonic orders, whereas the secondary plateau lies in the range of the 116th to 144th harmonic orders. 
The secondary plateau is about five orders of magnitude lower than the primary plateau. 
Our simulated spectrum confirms distinct features of solid HHG previously reported 
and  agrees  with other published  results~\cite{liu2017time, huang2017high, wu2015high, ikemachi2017trajectory, guan2016high}. 
 
\begin{figure}[h!]
\includegraphics[width=15 cm]{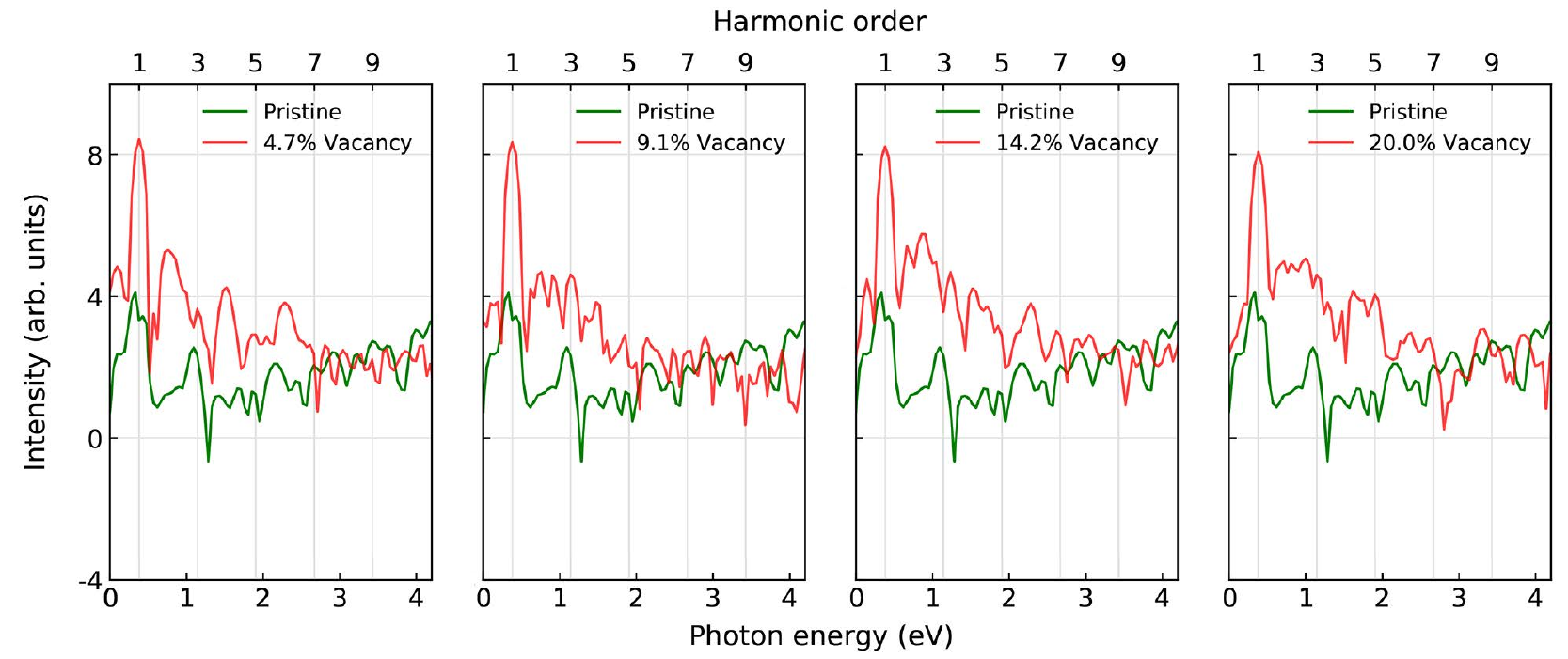}
\caption{Below band-gap harmonic spectra  for solids with vacancy concentrations 
(a) 4.7\%, (b) 9.1\%, (c) 14.2\%, and (d) 20.0\% (red line, light gray) compared to  
the spectrum of the pristine solid (green line, dark gray).} \label{fig3}
\end{figure}

The harmonic spectra of solids with vacancies are significantly different from the spectrum of a pristine solid,  as is evident from Fig.~\ref{fig2}. The spectrum corresponding to 4.7\% vacancy  
concentration is similar to the pristine one as the vacancy concentration is very low 
[see Fig.~\ref{fig2}(a)]. However, the total  energy cut off  increases significantly for  
9.1\%, 14.2\%, and 20.0\% vacancy concentrations as reflected in Figs.~\ref{fig2}(b), (c) 
and (d); respectively. 
Also, the spectrum profile of the harmonic yield gradually turns into a single slanted spectrum without a sharp distinction between the primary and secondary plateaus.
Not only does the total energy cut off increase  but also the below band-gap harmonics changes drastically for vacancy solids. There is an enhancement of 
approximately  two orders of magnitude 
in the below band-gap harmonic intensity  for solids with vacancies in comparison to the pristine solid,  
as reflected in Fig.~\ref{fig3}.

\begin{figure}[h!]
\includegraphics[width=15 cm]{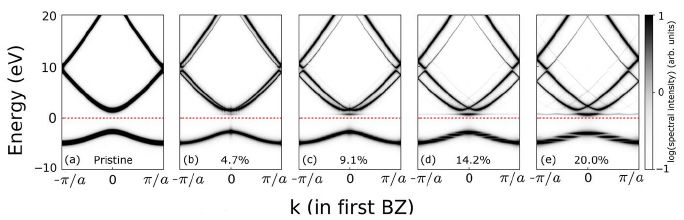}
\caption{Effective energy-band structures for (a) pristine solid and for solids with different vacancy concentrations 
(b) 4.7\%, (c) 9.1\%, (d) 14.2\%, and (e) 20.0\%. All the effective band structures are plotted using the same logarithmic spectral colormap. The red dashed line denotes the Fermi-energy level.}  \label{fig4}
\end{figure}

To understand the increase in energy cut off and spectral intensity enhancement in the below band-gap harmonics, the effective energy band structures of solids with different vacancy  concentrations are presented in Fig.~\ref{fig4}. 
The size of the  first Brillouin zone decreases as the size of the supercell increases. 
This means that the sizes of the first Brillouin zone, corresponding to solids with different vacancy  concentrations and to a pristine solid, are different. 
Note that different sizes of the supercell are 
used to realize different vacancy  concentrations. 
To compare the band structures of solids with different vacancy concentrations and a pristine 
solid for the identical size of the  first Brillouin zone, 
a band unfolding approach is used. 
Following the method given in Refs.~\citep{ku2010unfolding, popescu2012extracting, mayo2018band}, the energy-band structures of  solids with different vacancy concentrations are unfolded into the area of first Brillouin zone of a pristine solid. 

The uppermost valence and three lower conduction bands are the most affected bands due to  vacancy defects in solids, as evident from Fig.~\ref{fig4}.  
The inclusion of vacancy defects breaks the translational symmetry of the unit cell locally, although translational symmetry of the  supercell remains preserved globally. The breaking of the translational symmetry leads to splitting of the pristine energy bands and new defect states emerge~\cite{zhong2016first}.
As the vacancy concentration increases, new defect states become more prominent 
near the Fermi level. The splittings of  energy bands are predominantly observed in the conduction band region.

\begin{figure}[h!]
\includegraphics[scale=0.9]{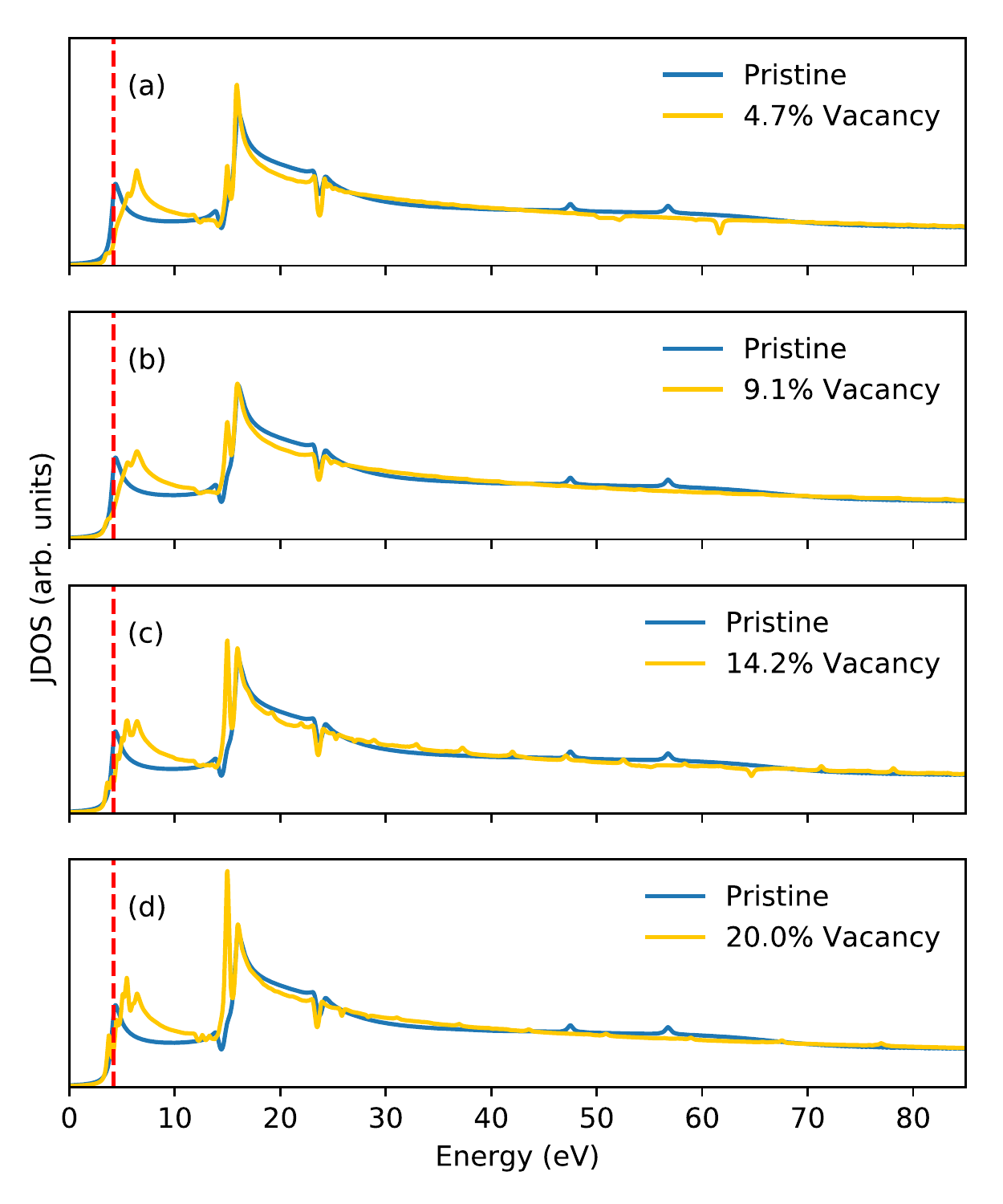}
\caption{Joint density of states (JDOS) for  different vacancy concentrations (a) 4.7\%, (b) 9.1 \%, (c) 14.2\%, and
(d) 20.0\%  in yellow color (light gray). 
The JDOS of the pristine is plotted as a reference in blue color (dark gray). 
The red dashed line represents the minimum band-gap energy between the uppermost valence band and the lowermost conduction band for a pristine solid.} \label{fig9}
\end{figure}

To better comprehend the total modification in the harmonic spectra, we analyze the joint density of states (JDOS) for different concentrations of vacancy defect. 
JDOS provides an estimate of possible transitions at a particular energy. From Fig.~\ref{fig9}, it is evident that the JDOS of a defect solid in the low-energy regime, upto 20 eV, 
is significantly modified in comparison to the JDOS of a pristine solid.  
There are additional JDOS in the below band-gap regime, accounting for the below band-gap enhancement in the harmonic spectra. The qualitative nature of the JDOS in the higher energy regime, above 25 eV, is also different for pristine and defect solids. 
These changes in the JDOS are responsible for the modulation in the harmonic spectra 
in the cutoff regime. Note that JDOS does not account for the probability of transitions, which may also affect the spectrum.

In a pristine solid, distinct interband coupling strengths and energy gaps between different pairs of conduction and valence bands lead to a different harmonic intensities in different plateaus. 
The new energy bands due to vacancy defects open new paths for the interband 
transitions, and these new interband transitions  
modify the electron dynamics.
As a result of these additional modifications, 
abrupt intensity drops between primary and secondary 
plateaus are reduced and the  plateaus gradually become indistinguishable  with the increase in the vacancy concentrations, as reflected in Fig~\ref{fig2}. 

\begin{figure}[h!]
\includegraphics[width= 18 cm]{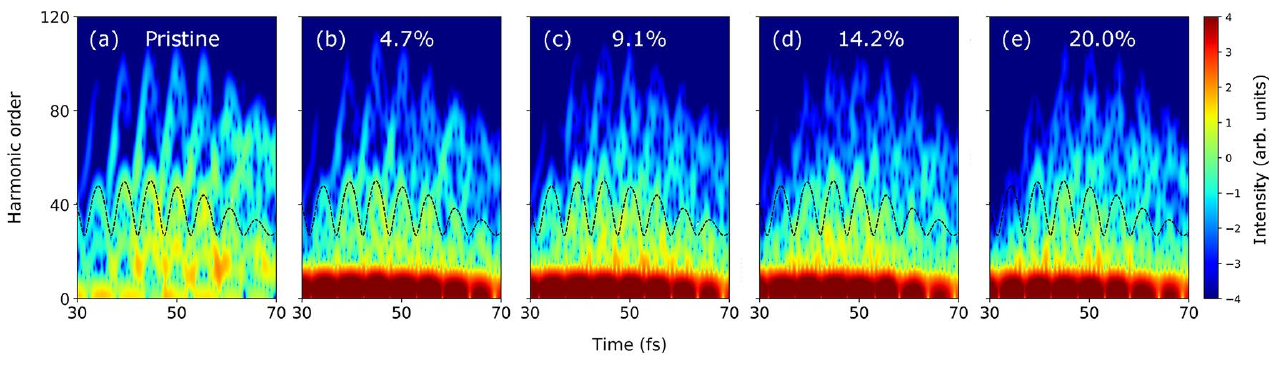}
\caption{Time-frequency maps corresponding to the harmonic spectra for  
(a) pristine solid and solids with vacancy concentrations of (b) 4.7\%, (c) 9.1\%, (d) 14.2\%, and (e) 20.0\%. The black dashed curve represents the well resolved trajectories of electrons between the second lowermost conduction band and uppermost valence band corresponding to the pristine solid.  $\sigma={1}/{(3\omega)}$ is used to perform the Gabor transform.} 
\label{fig5}
\end{figure}

Time-frequency maps corresponding to the harmonic spectra in Fig~\ref{fig2} 
are obtained by performing a Gabor transform and are presented in Fig~\ref{fig5}. 
The generation of electron-hole pairs and their trajectories  can be seen from the maps. 
The pristine solid has well-resolved states in the band structure, which translate 
into well-resolved trajectories in the time-frequency map [see Fig.~\ref{fig5}(a)], 
and  this is  consistent with earlier work as shown in Ref.~\cite{2016arXivdu}.  
Several additional  defect states in the band structure appear as the 
vacancy defects are  introduced  [see Figs.~\ref{fig4}(b)-(e)]. 
As a consequence of these additional states, 
additional paths for the electron-hole dynamics in a vacancy solid are possible and 
these paths  interfere with each other.  
The well-resolved trajectories in the time-frequency maps disappear as a result of additional 
interfering paths, as is visible in Figs.~\ref{fig5}(b)-(e). Similar ill-resolved trajectories  are 
discussed  in the time-frequency  map corresponding to HHG from an imperfect 
solid~\cite{Yu_imperfect_2019}.

The enhancement in the below band-gap harmonics in vacancy solid 
can be understood as follows: the defect states near the Fermi level open additional 
channels for interband transitions, which lead to an enhancement in the below band-gap harmonics~\cite{2019arXivmrudul}. The enhancement in the below band-gap harmonics is clearly visible in time-frequency maps [see red colored regions in 
Fig \ref{fig5}(b)-(e)]. 

\begin{figure}[h!]
\includegraphics[width= 15 cm]{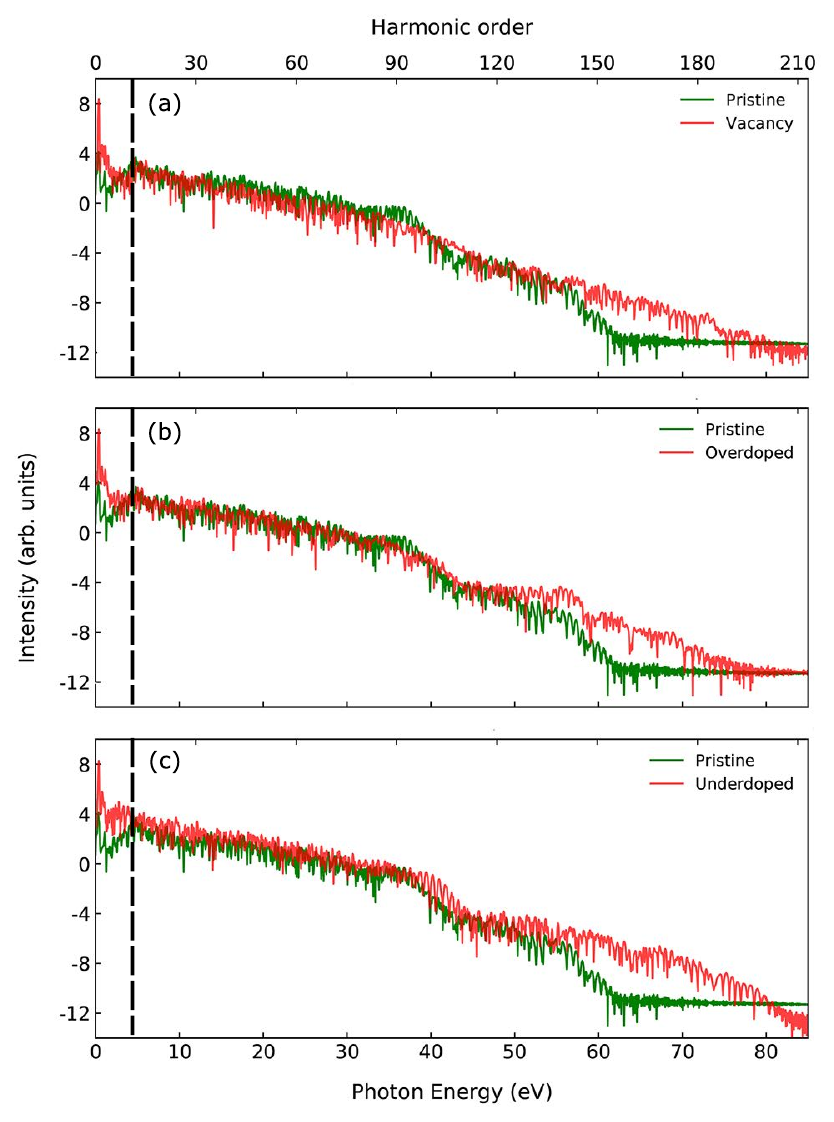}
\caption{High-order harmonic spectra for solids with different kinds of defects 
[red (light gray)]:  
(a) vacancies, (b) overdoping, and (c) underdoping and their comparison with the spectrum of  a pristine solid [green (dark gray)]. Here,  9.1\% defect concentration is used for different kinds of defects. The minimum band gap of 4.18 eV for a pristine solid is marked by the dashed black vertical line.} 
\label{fig6}
\end{figure}

\begin{figure}[h!]
\includegraphics[width=\linewidth]{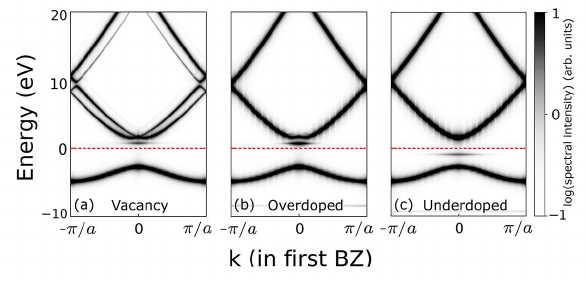}
\caption{Effective energy-band structures for (a) vacancy, (b) overdoping, and (c) underdoping defects with a fixed defect concentration of 9.1\%. All the effective band structures are plotted using the same logarithmic spectral colormap. The red dashed line denotes the Fermi-energy level.}
\label{fig7}
\end{figure}

To understand how HHG from a vacancy solid is different from HHG from solids with underdoping and overdoping defects, we present harmonic spectra 
from solids with these three kinds of
defects (vacancies, underdoping and overdoping, see Fig.~\ref{fig1}) and their 
comparison with the  pristine solid in Fig.~\ref{fig6}. 
Here, 9.1\% defect concentration is chosen for all  three cases. 
The harmonic spectrum for vacancy solid exhibits a slanted nature 
without having an abrupt variation [Fig.~\ref{fig6}(a)], whereas the spectra from underdoped 
or  overdoped solids show sharp transitions in intensity between the two plateaus 
[Figs.~\ref{fig6}(b) and (c)], much like HHG from the pristine solid. 
The significant variations among the spectra from the  three kinds of defects can be understood in terms of their effective band-structure as presented in Fig.~\ref{fig7}. 
There are new defect states near the Fermi level for overdoped and underdoped solids. 
In the case of vacancy defects, 
there are additional  defect states in the conduction and valence bands,  
in addition to the defect states near the Fermi level.  
 
\section{Conclusion}
In conclusion, the impact of different concentrations of vacancy defects in solid HHG is investigated. It is found that the total energy cutoff of the harmonic spectra increases with the increment in vacancy concentration. The intensity of below band-gap harmonics in solids 
with vacancy defects is  enhanced significantly in comparison to the harmonics in 
a pristine solid. The changes in the harmonic spectra of solids with vacancies are understood in terms of their effective unfolded band structures. The presence of a vacancy breaks the translational symmetry of the unit cell locally, which results in the splitting of pristine energy bands. 
The appearance of the new defect states opens additional paths for interband and intraband 
currents. 
This leads to the energy cutoff and harmonic yield different from the pristine solid. The time-frequency analysis of the HHG spectra confirms the enhancement in the below band-gap harmonics. It also helps us to conclude that the electron trajectories become ill-resolved with the increment in the vacancy concentration due to the  appearance of additional conduction bands and reduce the abrupt drop in the intensity of the harmonic yield between the plateaus. The single slanted harmonic spectrum carries a unique signature of vacancy defects which differs from the HHG from 
solids with underdoping or overdoping defects. 
The reduction in the abrupt drop in HHG intensity due to the band splitting makes HHG from vacancy defects distinctive.   
Our current work indicates that solids with  vacancy defects are potential band-gap materials 
for generation of high-energy radiation in XUV and soft x-ray energy regimes.  
HHG from solids with vacancy defects will be fascinating avenues for future work.

G.D. acknowledges support via a Ramanujan fellowship (SB/S2/ RJN-152/2015).  We are thankful 
for a fruitful discussion with Dr. Vladislav Yakovlev from the Max-Planck institute for Quantum Optics, Garching, Germany. 

\appendix
\section{Absorption Spectrum }\label{app:1}

\begin{figure}[h!]
\includegraphics[width=\linewidth]{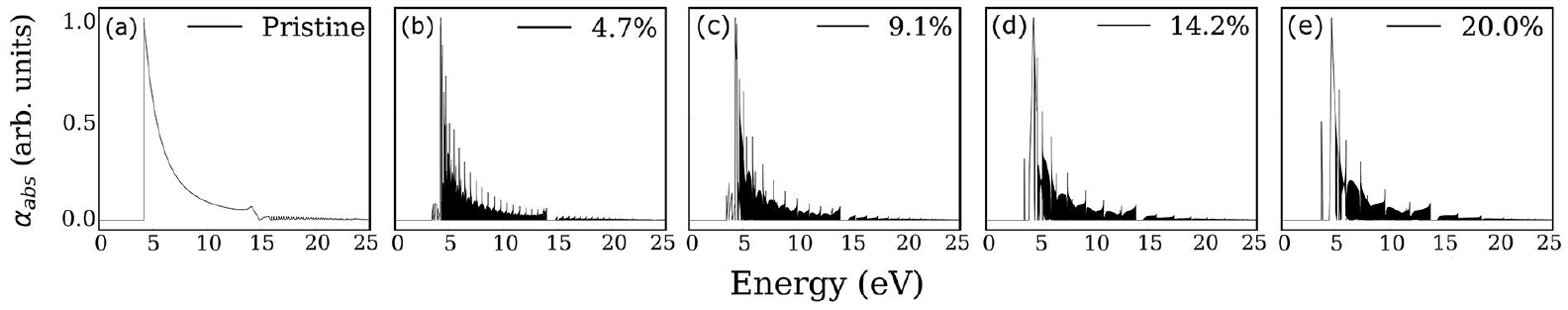}
\caption{Absorption spectra for (a) pristine solid and solids with different vacancy concentrations 
(b) 4.7\%, (c) 9.1\%, (d) 14.2\%, and (e) 20.0\%.} \label{fig8}
\end{figure}

Figure~\ref{fig8} present absorption spectra $\alpha_{abs}$ as a function of vacancy concentration. 
The modifications in the band structure due to vacancy defects  are captured by absorption spectra in the perturbative regime. 
The peak at 4.18 eV reflects the minimum band gap for a pristine solid [see Fig.~\ref{fig8}(a)]. Figures ~\ref{fig8}(b)-(e) show the emergence of multiple sharp peaks in the absorption spectra for different vacancy concentrations, which become apparent due to the appearance of new defect states in the energy-band structure.  The absorption peaks in the below-band-gap regime  indicate the transitions including defect states. These results support the findings regarding the enhancement of the harmonics near the band-gap regime.

%\bibliography{solid_HHG} 

\end{document}